\documentclass[namedreferences]{solarphysics}

\usepackage[optionalrh,solaromanenum,natbib,hyperref]{spr-sola-addons}
\usepackage{graphicx} 
\usepackage{color}    
\usepackage{breakurl} 
\hypersetup{colorlinks, citecolor=blue, filecolor=black, linkcolor=black, urlcolor=black}

\makeatletter

\newcommand{\Rmnum}[1]{\expandafter\@slowromancap\romannumeral #1@}
\makeatother

\newcommand{\degr}{\hbox{$^{\circ}$}}

\begin{document}

\begin{article}

\begin{opening}

\title{Performance of Major Flare Watches from the Max Millennium Program (2001\,--\,2010)}

\author{D.~S.~\surname{Bloomfield}$^{1}$\sep
        P.~T.~\surname{Gallagher}$^{1}$\sep
        W.~H.~\surname{Marquette}$^{2}$\sep
        R.~O.~\surname{Milligan}$^{3,4,5}$\sep
        R.~C.~\surname{Canfield}$^{6}$
       }

\runningauthor{D.S. Bloomfield {\it et al.}}
\runningtitle{Max Millennium Major Flare Watch Performance (2001\,--\,2010)}

\institute{$^{1}$ School of Physics, Trinity College Dublin, Dublin 2, Ireland\\
email: \url{shaun.bloomfield@tcd.ie} \\
           $^{2}$ Helio Research, 5212 Maryland Avenue, La Crescenta, CA 91214, USA\\
           $^{3}$ Astrophysics Research Centre, School of Mathematics and Physics, Queen's University Belfast, University Road, Belfast BT7 1NN, UK\\
           $^{4}$ Solar Physics Laboratory (Code 671), Heliophysics Science Division, NASA Goddard Space Flight Center, Greenbelt, MD 20771, USA\\
           $^{5}$ Department of Physics, Catholic University of America, 620 Michigan Avenue, N.E., Washington, DC 20064, USA\\
           $^{6}$ Montana State University, Bozeman, MT 59715, USA
           }

\begin{abstract}
The physical processes that trigger solar flares are not well understood and significant debate remains around processes governing particle acceleration, energy partition, and particle and energy transport. Observations at high resolution in energy, time, and space are required in multiple energy ranges over the whole course of many flares in order to build an understanding of these processes. Obtaining high-quality, co-temporal data from ground- and space- based instruments is crucial to achieving this goal and was the primary motivation for starting the Max Millennium program and Major Flare Watch (MFW) alerts, aimed at coordinating observations of all flares $\geqslant$\,X1 GOES X-ray classification (including those partially occulted by the limb). We present a review of the performance of MFWs from 1 February 2001 to 31 May 2010, inclusive, that finds: (1) 220 MFWs were issued in 3\,407\,days considered (6.5\% duty cycle), with these occurring in 32 uninterrupted periods that typically last 2\,--\,8\,days; (2) 56\% of flares $\geqslant$\,X1 were caught, occurring in 19\% of MFW days; (3) MFW periods ended at suitable times, but substantial gain could have been achieved in percentage of flares caught if periods had started 24\,h earlier; (4) MFWs successfully forecast X-class flares with a true skill statistic (TSS) verification metric score of 0.500, that is comparable to a categorical flare/no-flare interpretation of the NOAA Space Weather Prediction Centre probabilistic forecasts (TSS = 0.488).
\end{abstract}


\end{opening}

\section{Introduction}\label{s:intro} 

The Max Millennium program of solar flare research started on 14 June 1999 to facilitate the coordination of observing campaigns involving ground- and space- based instruments, particularly those with a limited ({\it i.e.}, not full-disk) field-of-view (FOV). The current form of the Max Millennium daily emails commenced on 1 February 2001, in preparation of supporting the NASA small explorer \emph{Reuven Ramaty High Energy Solar Spectroscopic Imager} \citep[RHESSI;][]{Lin:2002} X-ray mission that subsequently launched on 5 February 2002.

The scientific goal of the Max Millennium program is the same as that of RHESSI -- to understand impulsive energy release, particle acceleration, and both particle and energy transport in solar flares. Key questions are:
\begin{enumerate}
\item What role do high-energy particles play in the energy release process?
\item Do high-energy particles carry a significant fraction of the released energy?
\item What mechanisms accelerate both electrons and ions to high energies so efficiently?
\item What is the environment in which this energy release occurs?
\item What mechanisms transport the flare energy, the energetic particle component in particular, away from the energy release site?
\item What are the characteristic radiation signatures of flares that have potentially hazardous effects, and how do these flares occur and evolve?
\item What is the relationship between flares and coronal mass ejections (CMEs)?
\end{enumerate}
To answer these questions requires data of high resolution in energy, time, and space observed in multiple energy ranges and at multiple heights within the solar atmosphere. For example, the required data include:
\begin{itemize}
\item Hard X-ray, microwave, and mm-wave imaging spectroscopy to provide unique measurements of thermal and non-thermal electron parameters and pre-flare coronal magnetic fields \citep{White:2011, Krucker:2013};
\item Vector magnetograms to show the role of evolving magnetic field, via either emerging flux or reorienting of field vectors \citep{Murray:2012}. With Doppler data, they can be used to estimate the Poynting flux of magnetic energy into the corona and as boundary conditions for dynamic models of coronal field \citep{Kazachenko:2014};
\item Multi-band optical imaging to provide information on energy release, since optical continuum emission is thought to dominate the radiative energy budget of flares \citep{Kretzschmar:2011}.
\end{itemize}
No single instrument can provide all of these requirements and, given the limited FOVs of many high-cadence ground-based observatories, it is clearly crucial to obtain co-temporal data from ground- and space- based instruments.

The Max Millennium program facilitates this coordination through one of the Max Millennium chief observers (MMCOs) sending emails at least once per day that indicate a choice from one of the Max Millennium Observing Plans\footnote{\url{http://solar.physics.montana.edu/max_millennium/ops/observing.shtml}}. Nominally, these identify the solar active region most likely to produce flaring activity in the following 24-h period. When only low- to moderate- magnitude flares are likely, the daily email falls within the remit of Max Millennium Observing Plan (OP) 009 \emph{Default RHESSI Collaboration}. However, when high-magnitude flares are expected a Major Flare Watch (MFW) is called under Max Millennium OP 003 \emph{Region Likely to Produce Major Flares}. In this sense, MFW emails are a follow-on from the BEARALERTS service \citep{Zirin:1991} that was previously provided by Big Bear Solar Observatory (BBSO). An off-shoot from early MMCO activities was the creation of \url{http://www.solarmonitor.org} \citep[formerly the BBSO Active Region Monitor;][]{Gallagher:2002} that was designed to aid MMCOs in their target selection by displaying near real-time solar data from a variety of ground- and space-based observatories.

In this paper, we illustrate the efforts and performance of the Max Millennium program in coordinating observations of major solar flares from 1 February 2001 to 31 May 2010, inclusive. This is achieved by defining the criteria used to call MFWs and presenting their number and duration in Section~\ref{s:mfw}. The number and percentage of flares caught by MFWs and the percentage of MFW targets catching flares are given in Section~\ref{s:res}, while in Section~\ref{s:for_ver} forecast verification metrics are introduced and discussed. In Section~\ref{s:future} we describe changes to the program and future activities, while in Section~\ref{s:conc} we summarize the MFW results.

\section{Major Flare Watches (MFWs)}\label{s:mfw}

The specific goal of MFWs is to obtain multi-wavelength spectroscopic/imaging observations before, during, and after large flares. These highly-energetic events lead to high-quality RHESSI X-ray images and spectra that have high diagnostic potential and uniquely complement other instrument observations, as outlined in Section~\ref{s:intro}. For the period considered here, the Max Millennium program used the definition of a major flare as \emph{Geostationary Operational Environmental Satellite} (GOES) class X1 or greater, including those partially occulted by the limb.
 
\subsection{MFW Criteria}\label{ss:mfw_cri}

The definition of Max Millennium OP 003 \emph{Region Likely to Produce Major Flares} outlined below describes the circumstances under which an MMCO would have declared a MFW from 1 February 2001 to 31 May 2010\footnote{See \url{http://solar.physics.montana.edu/max_millennium/ops/op003/op003.html} for the currently operating MFW criteria.}. The MMCOs chose target active regions for OP 003 when \emph{any} of the following criteria were met:
\begin{enumerate}
\item A major flare ({\it i.e.}, $\geqslant$\,X1) has already occurred;
\item Large island $\delta$ configuration (from the \opencite{Hale:1919}, Mt. Wilson magnetic classification scheme that was extended by \citet{Kunzel:1960}) with sunspot area $\geqslant$\,500\,millionths of hemisphere ($\mu$H) -- even better if the region is reversed polarity. Bright H$\alpha$ emission present along the polarity inversion line;
\item Large $\delta$ configuration with bright H$\alpha$ plage (sunspot area $\geqslant$\,500\,$\mu$H) -- better still if reversed polarity. Bright H$\alpha$ emission along polarity inversion line;
\item Large elongated umbrae in pairs of opposite polarity (sunspot area $\geqslant$\,500\,$\mu$H), even if umbrae are not a $\delta$ configuration, so long as transverse magnetograms reveal sufficiently strong shear and strong-field inversion line length;
\item Emerging flux region coming up within a large existing sunspot group, if the emerging leading spots are adjacent to the existing trailing spots or vice-versa;
\item Rapidly moving large sunspot(s) moving either towards or into an opposite polarity sunspot.
\end{enumerate}
The scientific rationale behind these criteria are based on flare persistence \citep[item i: if a region produced one big flare, it will probably produce at least one more;][]{Sawyer:1986}, static signatures of stored free magnetic energy \citep[items ii\,--\,iv: large horizontal magnetic field gradients being an indicator of non-potential coronal magnetic field;][]{Schrijver:2005,Schrijver:2007}, rapidly changing magnetic topology \citep[item v: emerging magnetic fields creating magnetic null points -- potential sites for reconnection -- below existing overlying field;][]{MacTaggart:2011}, and dynamic signatures of increasing free magnetic energy \citep[item vi: plasma flows causing photospheric field to ramp up or shear;][]{Hudson:2008,Murray:2012}.

\subsection{MFW Occurrence}\label{ss:mfw_occ}

The first characteristic that we present is the number of individual MFW days, which is of particular interest to the chief observers and operators of reduced FOV instruments (both ground- and space- based) because this indicates the base-level observing time cost to responding to MFW alerts. In the 9.33 years considered here ({\it i.e.}, 3\,407\,days from 1 February 2001 to 31 May 2010, inclusive) only 220 MFW days were called by the MMCO team\footnote{Despite being sent, the MMCO email for 21 August 2002 is missing from the archive located here \url{http://solar.physics.montana.edu/hypermail/mmmotd/index.html}. This day was treated as having a MFW called on the same target region as the MFW days preceding and following (NOAA 10069), with an issue time of 13:00\,UTC (roughly the same as the previous day).}. This corresponds to just 6.5\% of the available observing days, indicating a particularly low observing overhead for any observatories following the MFW targets.

\begin{figure*}[!t]
\begin{center}
\includegraphics[keepaspectratio,width=\textwidth,trim=0 0 0 0]{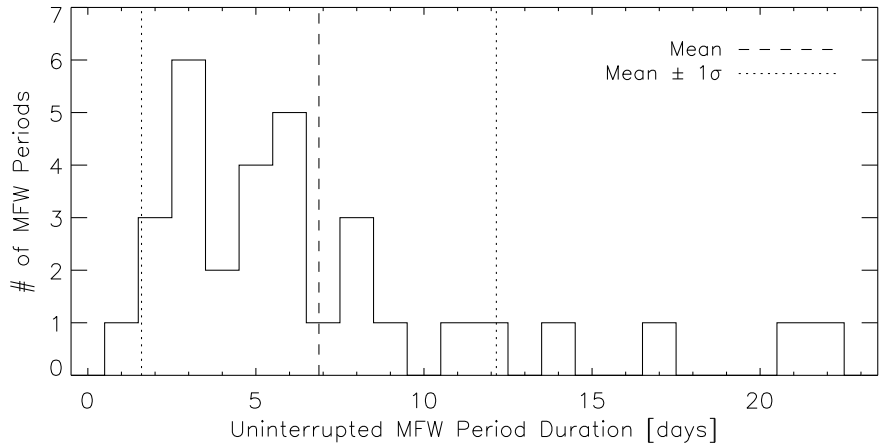}
\end{center}
\caption{Distribution of the total duration of uninterrupted MFW periods ({\it i.e.}, contiguous 24\,h MFW intervals) from 1 February 2001 to 31 May 2010, inclusive. The mean (6.9\,days) is indicated as a vertical dashed line, with the mean\,$\pm$\,1$\sigma$ (1.6 and 12.2\,days) as dotted lines.}
\label{fig:mfw_dur}
\end{figure*}

It is unlikely that individual instruments or observing facilities would be able to point to a MFW target region very soon after an MMCO email is sent. Hence, the duration of uninterrupted MFW periods ({\it i.e.}, the number of contiguous 24\,h MFW intervals) becomes an important characteristic. The 220 individual MFW days occur in 32 uninterrupted MFW periods, with the distribution of durations displayed in Figure~\ref{fig:mfw_dur}. This plot shows that uninterrupted MFW period durations are not normally distributed, while they have a mean\,$\pm$\,1$\sigma$ of 6.9\,$\pm$\,5.3\,days. Despite the non-normal nature of the distribution in Figure~\ref{fig:mfw_dur}, it is reasonable to state that ``typical'' MFW periods last for somewhere in the range 2\,--\,8\,days (covering 75\% of the 32 periods). This is encouraging for satellite chief observers for whom the time scale to upload new observation plans can be 1\,--\,2\,days, as these ``typical'' MFW period durations indicate that a target region will usually be able to be observed during an active MFW period despite systematic delays in responding to their initiation ({\it i.e.}, 88\% of the 32 periods last $>$\,2\,days).

It is worth noting that three MFW periods lasted longer than the $\approx$\,14\,days disk passage of an active region, and these involve the most recognisable NOAA region numbers of the cycle. The 17-day period corresponds to the 2003 Halloween storm regions (NOAA 10484 and 10486, separated by $\approx$\,70\degr\ in longitude), while the longest 2 periods correspond to pairs of flare-productive regions located at near-opposite longitudes (22\,days for NOAA 9393 and 9415, separated by $\approx$\,155\degr; 21\,days for NOAA 10030 and 10039, separated by $\approx$\,165\degr).

\section{Results and Discussion}\label{s:res}

Here we present simple measures of success in terms of the number (Section~\ref{ss:res_num_fla}) and percentage (Section~\ref{ss:res_per_fla}) of flares caught by MFW targets, before highlighting the percentage of MFW targets that caught these flares (Section~\ref{ss:res_per_mfw}). We present four thought experiments showing how the percentage of flares caught varies with 24\,h adjustments to the start/end times of MFW periods (Section~\ref{ss:res_mfw_var}), before discussing the regions that flares were missed from (Section~\ref{ss:res_mis_fla}).

\subsection{Numbers of Flares Caught}\label{ss:res_num_fla}

\begin{figure*}[!th]
\begin{center}
\includegraphics[keepaspectratio,width=0.94\textwidth,trim=0 0 0 0]{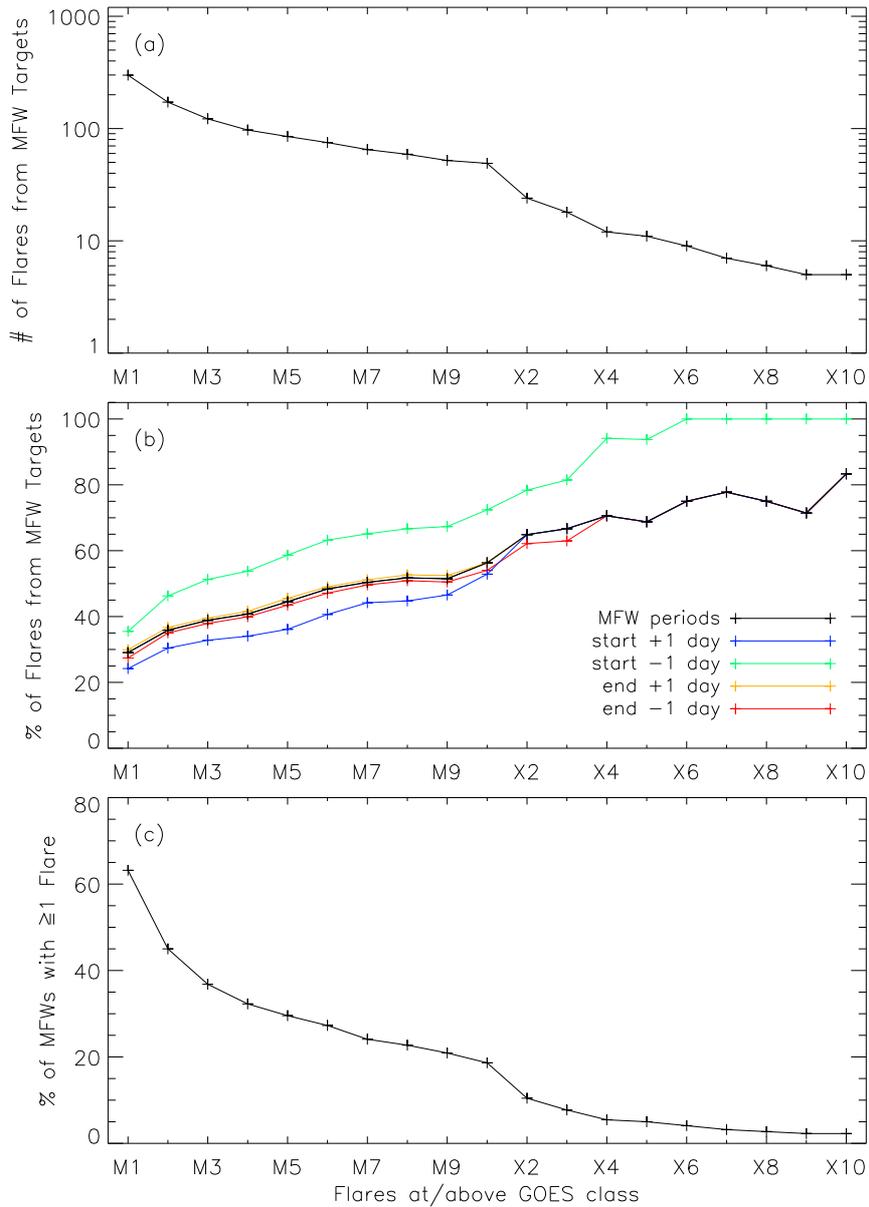}
\end{center}
\caption{(a) Number of flares at/above a chosen GOES class produced by a MFW target region during a MFW day ({\it i.e.}, within 24\,h of the MFW UTC email issue times) from 1 February 2001 to 31 May 2010, inclusive. (b) Percentage of flares at/above a chosen GOES class that were produced by a MFW target region during a MFW day, relative to the total number of flares from all regions in the entire time range. Different curves are shown for various 24-h offsets in the start and end times of uninterrupted MFW periods. (c) Percentage of MFW days whose target regions produced at least one flare at/above a chosen GOES class.}
\label{fig:fla_per_var}
\end{figure*}

The main characteristic for those parties interested in maximizing the scientific return from coordinated flare observations is the number of flares that occurred in MFW target regions. Although MFWs aim to catch flares $\geqslant$\,X1, Table~\ref{tbl:fla_per} and Figure~\ref{fig:fla_per_var}(a) present the number of flares as a function of the minimum GOES class considered. This shows the decrease in numbers with magnitude expected from the power-law occurrence of flares \citep{Hudson:1969} with a cumulative distribution power-law slope of $\beta_{\mathrm{P}} = 0.9$\,$\pm$\,0.4 that equates to a power-law slope of $\alpha_{\mathrm{P}} = \beta_{\mathrm{P}} + 1 = 1.9$\,$\pm$\,0.4, consistent with $\alpha_{\mathrm{P}} = 2.0$\,$\pm$\,0.1 found by \citet{Aschwanden:2012} when averaging over 1975\,--\,2011.

\begin{table}[!t]
\caption{Number and percentage of flares from MFW target regions, total number of flares occurring, and number and percentage of MFW targets producing at least one flare at/above a chosen GOES class (from 1 February 2001 to 31 May 2010, inclusive).}
\label{tbl:fla_per}
\begin{tabular}{lccccc}
\hline
GOES             & \multicolumn{3}{c}{Flares}                & \multicolumn{2}{c}{MFW targets (220 total)}\\
class            & \# caught by & whole-Sun   & \% caught by & \# producing         & \% producing\\
                 & MFW target   & total \#    & MFW target   & $\geqslant$\,1 flare & $\geqslant$\,1 flare\\
\hline
$\geqslant$\,M1  & 299          & 1\,029      & 29           & 139                  & 63\\
$\geqslant$\,M2  & 172          &    480      & 36           &  99                  & 45\\
$\geqslant$\,M3  & 122          &    314      & 39           &  81                  & 37\\
$\geqslant$\,M4  &  97          &    238      & 41           &  71                  & 32\\
$\geqslant$\,M5  &  85          &    191      & 45           &  65                  & 30\\
$\geqslant$\,M6  &  75          &    155      & 48           &  60                  & 27\\
$\geqslant$\,M7  &  65          &    129      & 50           &  53                  & 24\\
$\geqslant$\,M8  &  59          &    114      & 52           &  50                  & 23\\
$\geqslant$\,M9  &  52          &    101      & 51           &  46                  & 21\\
$\geqslant$\,X1  &  49          &     87      & 56           &  41                  & 19\\
$\geqslant$\,X2  &  24          &     37      & 65           &  23                  & 10\\
$\geqslant$\,X3  &  18          &     27      & 67           &  17                  &  8\\
$\geqslant$\,X4  &  12          &     17      & 71           &  12                  &  5\\
$\geqslant$\,X5  &  11          &     16      & 69           &  11                  &  5\\
$\geqslant$\,X6  &   9          &     12      & 75           &   9                  &  4\\
$\geqslant$\,X7  &   7          &      9      & 78           &   7                  &  3\\
$\geqslant$\,X8  &   6          &      8      & 75           &   6                  &  3\\
$\geqslant$\,X9  &   5          &      7      & 71           &   5                  &  2\\
$\geqslant$\,X10 &   5          &      6      & 83           &   5                  &  2\\
\hline
\end{tabular}
\end{table}

\subsection{Percentages of Flares Caught}\label{ss:res_per_fla}

Although Figure~\ref{fig:fla_per_var}(a) is interesting in terms of absolute numbers, it is more useful to consider the percentage of flares at/above a GOES class occurring in MFW target regions as a first-order measure of MFW success. This characteristic is shown in Table~\ref{tbl:fla_per} and the black curve in Figure~\ref{fig:fla_per_var}(b), exhibiting an upward trend with increasing flare magnitude. This tells us that a relatively larger fraction of high-magnitude flares are being produced from MFW targets than is the case for moderate-magnitude flares. To highlight some numbers in particular, MFW targets caught 29\% of flares $\geqslant$\,M1, 45\% $\geqslant$\,M5, 56\% $\geqslant$\,X1, and 69\% $\geqslant$\,X5.

\subsection{Percentages of MFW Targets Catching Flares}\label{ss:res_per_mfw}

In addition to the number and percentage of flares caught, the percentage of MFW targets catching flares is an important characteristic of their success. To study this, Table~\ref{tbl:fla_per} and Figure~\ref{fig:fla_per_var}(c) display the percentage of all 220 MFW days whose target region produced $\geqslant$\,1 flare, again as a function of minimum GOES class considered. It is worthwhile revisiting the percentages of flares caught to help put the performance of the MFW target region selection in context. Table~\ref{tbl:fla_per} shows us that 63\% of MFW targets produced flaring $\geqslant$\,M1, with 29\% of all flares above that level caught ({\it i.e.}, 2.2 $\geqslant$\,M1 flares produced on average over 139 MFW days). Similarly, 30\% of MFW targets produced flaring $\geqslant$\,M5, with 45\% of all flares above this level caught ({\it i.e.}, 1.3 $\geqslant$\,M5 flares produced on average over 65 MFW days). Of most interest here, due to the focus on the performance of MFWs, 19\% of target regions produced flaring $\geqslant$\,X1, with 56\% of all flares above this level caught ({\it i.e.}, 1.2 $\geqslant$\,X1 flares produced on average over 41 MFW days). Finally, 5\% of MFW targets produced flaring $\geqslant$\,X5, with 69\% of all flares above this level caught ({\it i.e.}, 1.0 $\geqslant$\,X5 flares produced on average over 11 MFW days).

\subsection{Variation of MFW Start and End Times}\label{ss:res_mfw_var}

It is a useful exercise at this point to consider the change in flare-catching success that would result from the uninterrupted MFW periods instead starting one day earlier or later or ending one day earlier or later. The percentage of flares that occurred from MFW targets for these four possibilities are included as additional colour curves in Figure~\ref{fig:fla_per_var}(b), with each described and discussed in detail below.
\begin{description}
\item[start +1 day] The first case, presented by the blue curve in Figure~\ref{fig:fla_per_var}(b), is that corresponding to a 24-h delay in responding to the first MFW issued in an uninterrupted MFW period. This mimics the scenario of scientific satellite operators who have an inherent delay period resulting from the time scale required to approve and upload new observation plans. Encouragingly, the percentage of flares $\geqslant$\,X2 is unaffected by such a 24-h delay and for flare magnitudes M1\,--\,X1 the percentage of flares missed is only 3\,--\,8\%.
\item[start $-$1 day] The second case, presented by the turquoise curve in Figure~\ref{fig:fla_per_var}(b), is that corresponding to all uninterrupted MFW periods beginning 24\,h earlier than the first (actually issued) MFW in each period. To determine the success of these extra MFW days the target region was copied from the first day in each period. This curve is the one that deviates the most from the true MFW ({\it i.e.}, black) curve, epitomizing the MMCO ``holy grail'' -- catching the first major flare from a region. The percentage of flares caught would have been systematically larger by typically 10\,--\,25\% and, probably most importantly, all flares $\geqslant$\,X6 magnitude would have been caught.
\item[end +1 day] The third case, presented by the yellow curve in Figure~\ref{fig:fla_per_var}(b), is that corresponding to a 24-h delay in ending an uninterrupted MFW period. Similar to before, to determine the success of these extra MFW days the target region was duplicated from that of the final MFW day in each period. This represents the situation that would have occurred if MMCOs had always issued an additional MFW day that extended each of the uninterrupted MFW periods. This curve lies encouragingly close to the true MFW ({\it i.e.}, black) curve, with only a marginal increase of $<$\,1\% for catching flare magnitudes $<$\,X1 and no change at all for $\geqslant$\,X1 flares.
\item[end $-$1 day] The fourth case, presented by the red curve in Figure~\ref{fig:fla_per_var}(b), is that corresponding to each uninterrupted MFW period ending 24\,h earlier than the final (actually issued) MFW in that period. This represents the scenario that would have occurred if MMCOs had terminated each MFW period one day early. Like the previous case, this curve lies close to the true MFW ({\it i.e.}, black) curve with a marginal decrease of $<$\,1\% for flare magnitudes $<$\,X1 and a slightly larger decrease of 2\,--\,4\% for flares in the range X1\,--\,X3.
\end{description}
The final two cases give strong indications that the MMCOs have terminated the uninterrupted MFW periods at suitable times -- {\it i.e.}, no real benefit seen when extending MFW periods by 24\,h, with only very minor loss of M1\,--\,M9 flares and some loss of X1\,--\,X3 flares being caught when ceasing MFW periods 24\,h early. However, the area of greatest potential for increasing the number/percentage of flares caught by MFW targets clearly lies in commencing MFW periods at earlier times. Starting each of the 32 MFW periods 24\,h earlier would have given 16\% more $\geqslant$\,X1 flares being caught ({\it i.e.}, an additional 14 events within 13 individual days). These cases highlight the reactive form that some MFWs have for being initiated, in that the first MFW criterion was the reason for issuing them ({\it i.e.}, a $\geqslant$\,X1 flare has already occurred). It should be noted though, that this only corresponds to 13 of the 32 uninterrupted MFW periods ({\it i.e.}, 41\%).

\subsection{Missed $\geqslant$\,X1 Flares}\label{ss:res_mis_fla}

The quantities given in Sections~\ref{ss:res_num_fla}\,--\,\ref{ss:res_per_mfw} indicate that MFWs were successful at catching flares $\geqslant$\,X1, with the target regions on 41 MFW days producing 49 flares out of 87. It is worth considering what may have led MMCOs to not call MFWs on the regions that produced the 38 missed $\geqslant$\,X1 flares, which are individually detailed in Table~\ref{tbl:mfw_fails}. First, knowing what flaring these regions ended up producing, they can be broken down into the following five categories:
\begin{description}
\item[Unknown source] This was the case for only 1 missed flare, a highly impulsive X1.1 on 25 November 2001 when context solar images were sparse in time;
\item[``One-shot wonder''] The largest category, corresponding to 16 of the missed flares. In terms of forecast verification (see Section~\ref{s:for_ver}), these contribute half of the incorrect ``No MFW'' days ({\it i.e.}, false negatives). However, only four of the 16 regions in this category subsequently had MFWs issued with them as target regions. This indicates that MMCOs were correct in determining that 75\% of these regions were unlikely to produce further $\geqslant$\,X1 flares;
\item[Future unsuccessful target] Regions producing multiple $\geqslant$\,X1 flares with unsuccessful MFWs issued. Of the 10 flares that were missed, three occurred during successful MFWs on other targets ({\it i.e.}, from NOAA 10484 and 10488 on MFW days for 10486) and two bracketed a MFW period that was too short ({\it i.e.}, a 3-day MFW period initiated on NOAA 10656 in reaction to an X1.0, while an X1.8 occurred just 4.5\,h after the MFW period ceased);
\item[Future successful target] 
These 10 flares resulted in MFWs being initiated reactively on eight regions, amounting to 25\% of all MFW periods (8/32);
\item[Past successful target] 
Only one region fell in this category (NOAA 10039), with one missed flare occurring 27\,h after its MFW period ended and shortly before completing west-limb transit.
\end{description}

\begin{table}[!t]
\caption{Details of $\geqslant$\,X1 flares missed by MFWs and their source active regions. Entries in parentheses indicate flares that occurred during MFWs on other targets (peak flux), regions first designated on later days (NOAA No.), and class additions on next UT day (Mt. Wilson).}
\label{tbl:mfw_fails}
\begin{tabular}{lccccc}
\hline
Missed flare category              & \multicolumn{2}{c}{GOES flare details} & \multicolumn{3}{c}{Region details (at 00:00\,UT)}\\
                                   & Start            & Peak                & NOAA     & HG       & Mt.\\
                                   & time             & flux                & No.      & location & Wilson\\
\hline
Unknown source\dotfill             & 2001-11-25T09:45 &  X1.1               & \ldots   & \ldots   & \ldots\\
\hline
``One-shot wonder''\dotfill        & 2001-06-23T04:02 &  X1.2               &  \ 9511  & N10E25   & $\beta\gamma$($\delta$)\\
\dotfill                           & 2001-08-25T16:23 &  X5.3               &  \ 9591  & S18E42   & $\beta\gamma\delta$\\
\dotfill                           & 2001-09-24T09:32 &  X2.6               &  \ 9632  & S18E32   & $\beta\gamma\delta$\\
\dotfill                           & 2001-11-04T16:03 &  X1.0               &  \ 9684  & N05W15   & $\beta\gamma$($\delta$)\\
\dotfill                           & 2001-12-28T20:02 &  X3.4               & \ (9767) & \ldots   & \ldots\\
\dotfill                           & 2002-04-21T00:43 &  X1.5               &  \ 9906  & S14W79   & $\beta\gamma$\\
\dotfill                           & 2002-05-20T15:21 &  X2.1               &  \ 9961  & S22E76   & $\beta$($\gamma$)\\
\dotfill                           & 2002-07-03T02:08 &  X1.5               &   10017  & S18W51   & $\beta\gamma$($\delta$)\\
\dotfill                           & 2002-08-30T12:47 &  X1.5               &   10095  & N07E76   & $\beta$($\gamma$)\\
\dotfill                           & 2002-10-31T16:47 &  X1.2               &  (10183) & \ldots   & \ldots\\
\dotfill                           & 2003-06-15T23:25 &  X1.3               &  (10386) & \ldots   & \ldots\\
\dotfill                           & 2004-02-26T01:50 &  X1.1               &   10564  & N14W14   & $\beta\gamma$($\delta$)\\
\dotfill                           & 2004-10-30T11:38 &  X1.2               &   10691  & N13W13   & $\beta$($\gamma$)\\
\dotfill                           & 2005-01-01T00:01 &  X1.7               &   10715  & N04E34   & $\beta\gamma\delta$\\
\dotfill                           & 2005-07-14T10:16 &  X1.2               &   10786  & N11W84   & $\beta\gamma\delta$\\
\dotfill                           & 2005-07-30T06:17 &  X1.3               &   10792  & N11E66   & $\beta\gamma$($\delta$)\\
\hline
Future unsuccessful target\dotfill & 2001-12-11T07:58 &  X2.8               &  \ 9733  & N14E44   & $\beta\gamma$\\
\dotfill                           & 2001-12-13T14:20 &  X6.2               &  \ 9733  & N14E18   & $\beta\gamma\delta$\\
\dotfill                           & 2003-03-17T18:50 &  X1.5               &   10314  & S14W26   & $\beta\gamma\delta$\\
\dotfill                           & 2003-03-18T11:51 &  X1.5               &   10314  & S16W39   & $\beta\gamma\delta$\\
\dotfill                           & 2003-10-19T16:29 &  X1.1               &   10484  & N05E68   & $\beta$($\gamma\delta$)\\
\dotfill                           & 2003-10-26T17:21 & (X1.2)              &   10484  & N04W28   & $\beta\gamma\delta$\\
\dotfill                           & 2003-11-03T01:09 & (X2.7)              &   10488  & N08W68   & $\beta\gamma\delta$\\
\dotfill                           & 2003-11-03T09:43 & (X3.9)              &   10488  & N08W68   & $\beta\gamma\delta$\\
\dotfill                           & 2004-08-13T18:07 &  X1.0               &   10656  & S13W09   & $\beta\gamma\delta$\\
\dotfill                           & 2004-08-18T17:29 &  X1.8               &   10656  & S14W77   & $\beta\gamma\delta$\\
\hline
Future successful target\dotfill   & 2001-04-02T10:58 & (X1.1)              & \ (9415) & \ldots   & \ldots\\
\dotfill                           & 2001-04-03T03:25 & (X1.2)              & \ (9415) & \ldots   & \ldots\\
\dotfill                           & 2001-10-22T17:44 &  X1.2               &  \ 9672  & S19E26   & $\beta$($\gamma\delta$)\\
\dotfill                           & 2002-07-20T21:04 & (X3.3)              &  (10039) & \ldots   & \ldots\\
\dotfill                           & 2003-05-27T22:56 &  X1.3               &   10365  & S06W06   & $\beta$($\gamma\delta$)\\
\dotfill                           & 2003-05-28T00:17 &  X3.6               &   10365  & S07W19   & $\beta\gamma\delta$\\
\dotfill                           & 2003-10-23T08:19 & (X5.4)              &   10486  & S16E81   & $\alpha$($\beta\gamma\delta$)\\
\dotfill                           & 2004-07-15T01:30 & (X1.8)              &   10649  & S10E53   & $\beta\gamma\delta$\\
\dotfill                           & 2005-09-07T17:17 &  X17.0              &  (10808) & \ldots   & \ldots\\
\dotfill                           & 2006-12-05T10:18 &  X9.0               &  (10930) & \ldots   & \ldots\\
\hline
Past successful target\dotfill     & 2002-08-03T18:59 &  X1.0               &   10039  & S15W70   & $\beta\gamma$\\
\hline
\end{tabular}
\end{table}

Second, it is worth considering the spatial locations of these regions at the time MMCOs were considering whether or not to issue MFWs. Of the 38 missed $\geqslant$\,X1 flares, 11 originated in regions beyond the east limb or at heliographic longitudes between E90 and E65 (notably five of which were in the ``one shot wonders'' category discussed above). These correspond to regions with no known magnetic classification or one suffering from significant foreshortening effects that mask internal spot polarity mixing and $\delta$ configurations. All 11 of these regions were designated as $\beta\gamma$ or $\beta\gamma\delta$ when located at E65\,--\,E60, by which point their magnetic complexity is unambiguously identified. The only way that MMCOs and other forecasters could have had the necessary knowledge of the true magnetic complexity and prior flaring history of these active regions located close to or beyond the east limb (as viewed from Earth) would be through a solar observing satellite positioned at an L5 vantage point. If such a mission had been active in the time range considered here, the success rate of MFWs catching $\geqslant$\,X1 flares could potentially have increased from 56\% to 69\%.

\section{Forecast Verification}\label{s:for_ver}

In comparison to the simple characteristics of absolute number and percentage of flares caught by the MFW target regions, formal forecast verification metrics are aimed at those parties interested in understanding or quantifying both the successes and failures of the daily MMCO emails. Forecast verification takes into account all daily forecasts, encompassing both the nominal MMCO email messages ({\it i.e.}, forecasts of $<$\,X1 activity) and the much less frequent MFW messages ({\it i.e.}, forecasts of $\geqslant$\,X1 activity). The issuing of daily emails that either call a MFW or not leads to a categorical ({\it i.e.}, flare/no-flare) forecast, which can then be evaluated with the posterior knowledge of whether or not flaring $\geqslant$\,X1 occurred from the specified MFW target region in the following 24-h period.

The combination of categorical forecasts and categorical flare occurrence leads to a $2\times2$ contingency table, such as that presented in Table~\ref{tbl:mfw_ct}. Each individual day considered here therefore falls into one of the following four possibilities,
\begin{itemize}
\item true positive (TP), MFW issued and at least one $\geqslant$\,X1 flare occurred,
\item false positive (FP), MFW issued and no $\geqslant$\,X1 flares occurred,
\item false negative (FN), no MFW issued and at least one $\geqslant$\,X1 flare occurred,
\item true negative (TN), no MFW issued and no $\geqslant$\,X1 flares occurred,
\end{itemize}
with these labels presented in parentheses in Table~\ref{tbl:mfw_ct}. The sum of all contingency table elements, $N = \mathrm{TP + FP + FN + TN}$, then corresponds to the total number of days in the forecasting interval ({\it i.e.}, $N=3\,407$ from 1 February 2001 to 31 May 2010, inclusive).

\begin{table}[!t]
\caption{Forecast contingency table for MFWs from 1 February 2001 to 31 May 2010, inclusive.}
\label{tbl:mfw_ct}
\begin{tabular}{lrr}
\hline
Observed       & \multicolumn{2}{c}{Forecasted}\\
flaring        & MFW      & No MFW\\
\hline
$\geqslant$ X1 &  41 (TP) &     33 (FN)\\
        $<$ X1 & 179 (FP) & 3\,154 (TN)\\
\hline
\end{tabular}
\end{table}

The four contingency table elements can be combined in a wide variety of ways to form numerous verification metrics and skill scores ({\it i.e.}, metrics measured relative to some reference forecast). The best metrics typically penalize forecasts with high numbers of false alarms (FP), missed flares (FN), or a combination of the two. However, not all metrics take these quantities into account. A selection of the most commonly used metrics are briefly introduced in Section~\ref{ss:ver_met_sss} before being applied in Section~\ref{ss:mfw_met_sss} to evaluate the MFWs. Finally, in Section~\ref{ss:oth_met_sss} we compare the MFW metrics with other $\geqslant$\,X1 flare forecasting efforts.

\subsection{Verification Metrics and Skill Scores}\label{ss:ver_met_sss}

Accuracy (ACC, also known as percentage correct) defines the fraction of all forecasts that are correct in either sense ({\it i.e.}, either a MFW was issued and $\geqslant$\,X1 events were observed or no MFW was issued and no $\geqslant$\,X1 events were observed),
\begin{equation}\label{eqn:acc}
\mathrm{ACC} = \frac{\mathrm{TP + TN}}{N} \ ,
\end{equation}
but does not penalize for incorrect forecasts in the form of non-MFW $\geqslant$\,X1 days (FN) or false alarm MFWs (FP). The latter of these forms of incorrect forecasts are represented by the false alarm ratio (FAR),
\begin{equation}\label{eqn:far}
\mathrm{FAR} = \mathrm{\frac{FP}{TP + FP}} \ .
\end{equation}

Probability of detection (POD) is useful for interpreting the success achieved in forecasting the days where an event does happen, as it defines the fraction of $\geqslant$\,X1 days that were correctly forecast as MFW days,
\begin{equation}\label{eqn:pod}
\mathrm{POD} = \mathrm{\frac{TP}{TP + FN}} \ .
\end{equation}
It should be noted that POD uses only the upper half of the contingency table, while a related metric that uses only the lower half is the probability of false detection (POFD). This metric is useful for interpreting the failure achieved in forecasting the days where no event happens, as it defines the fraction of $<$\,X1 days that were incorrectly forecast as MFW days,
\begin{equation}\label{eqn:pofd}
\mathrm{POFD} = \mathrm{\frac{FP}{FP + TN}} \ .
\end{equation}
It can be easily seen that together POD and POFD use all of the contingency table elements and they are readily combined to achieve the \citet{Hanssen:1965} discriminant (also known as Peirce's skill score or the true skill statistic; TSS),
\begin{equation}\label{eqn:tss}
\mathrm{TSS} = \mathrm{POD} - \mathrm{POFD} \ .
\end{equation}
This metric is sensitive to all forms of correct and incorrect forecast because it uses all four contingency table elements and, most importantly, it is insensitive to the underlying climatology ({\it i.e.}, for this work, the relative frequency of $\geqslant$\,X1 days to $<$\,X1 days). This is due to POD and POFD being separately calculated from the upper and lower halves of the contingency table, respectively. These reasons lead to TSS being the preferred metric for intercomparison of forecast method performance over different time ranges \citep{Bloomfield:2012}.

Another metric that uses all four elements is the \citet{Heidke:1926} skill score (HSS) that assesses the forecast accuracy relative to random chance. This is calculated by removing the expected number of correct forecasts due to random chance, $E_{\mathrm{rand}}$, from Equation~(\ref{eqn:acc}),
\begin{equation}\label{eqn:hss}
\mathrm{HSS} = \frac{\mathrm{TP + TN} - E_{\mathrm{rand}}}{N - E_{\mathrm{rand}}} \ ,
\end{equation}
where,
\begin{equation}\label{eqn:hss_exp_ran}
E_{\mathrm{rand}} = \frac{\mathrm{(TP + FN)(TP + FP) + (TN + FN)(TN + FP)}}{N} \ ,
\end{equation}
and the random-chance contributions to both correct forecasts of TP (first term) and TN (second term) are accounted for. However, unlike TSS, the HSS metric is highly sensitive to the event climatology and so is difficult to use in comparing different forecast methods (or even the same method) over different time ranges \citep{Bloomfield:2012}.

The metric of forecast bias (BIAS) does not use the entire contingency table (notably excluding TN that dominates because of the rare nature of $\geqslant$\,X1 flares), but is still useful because it defines the relative frequency of the forecasted MFW days to the observed $\geqslant$\,X1 days,
\begin{equation}\label{eqn:bias}
\mathrm{BIAS} = \mathrm{\frac{TP + FP}{TP + FN}} \ .
\end{equation}
For this metric, values below unity denote underforecasting ({\it i.e.}, less forecasted MFW days than observed $\geqslant$\,X1 days) and values above denote overforecasting ({\it i.e.}, more forecasted MFW days than observed $\geqslant$\,X1 days).

\subsection{MFW Forecast Performance}\label{ss:mfw_met_sss}

Combining the forecast contingency table values in Table~\ref{tbl:mfw_ct} with Equations~(\ref{eqn:acc})\,--\,(\ref{eqn:bias}) results in the values presented in column 9 of Table~\ref{tbl:mfw_sss} that, for reference, includes the allowed value range (column 2) and perfect forecast value (column 3) for each of the metrics/skill scores described above. It can be seen that MFWs have an apparently high ACC of 0.938, but this metric is misleading for forecasts of rare events (like $\geqslant$\,X1 here). To be an extremely good forecast the ACC does not only need to approach 1, but should aim to be on the order of the climatology of the dominant observation. For this work that is the relative occurrence of $<$\,X1 days -- {\it i.e.}, (FP + TN)/$N$ = 3\,333/3\,407 = 0.978. This corresponds to the ACC that would have been achieved by MMCOs \emph{never} issuing a MFW -- a scenario involving no skill whatsoever and incapable of ever catching $\geqslant$\,X1 days.

The decrease in ACC to below that of the $<$\,X1 climatology is a direct result of MMCOs actively attempting to forecast $\geqslant$\,X1 flaring days, which also leads to a non-zero FAR (since never issuing a MFW would result in FP, and hence FAR, being zero). The FAR value of 0.814 found here for MFWs is quite high. However, this needs to be taken in the context of resulting from only a small portion (179 incorrect MFW days from a total of 220) of all 3\,407 days, since the dominant TN value is absent from this metric. Large FAR is to be expected when the BIAS that is achieved is considered, with Table~\ref{tbl:mfw_sss} showing that MMCOs forecast almost three times as many MFW days as there were $\geqslant$\,X1 days observed. However, it is again worth recalling that BIAS does not take into account the dominant TN value.

\begin{table*}
\caption{Forecast verification metrics and skill scores for MFWs (from 1 February 2001 to 31 May 2010, inclusive) and other published $\geqslant$\,X1 flare forecasts (over different time periods).}
\label{tbl:mfw_sss}
\begin{tabular}{lccccccccc}
\hline
Metric  & Allowed                      & Perfect  & \multicolumn{5}{c}{Published values}        & MFW\\
/ Skill & value                        & forecast & Colak  & Song   & Mason  & B'field & Crown  & value\\
score   & range                        & value    & (2009) & (2009) & (2010) & (2012)  & (2012) &\\
\hline	
ACC     & ~~~0 $\rightarrow$ 1~        & 1        & 0.981  & 0.945  &  0.694 &  0.881  & 0.996 & 0.938\\
FAR     & ~~~0 $\rightarrow$ 1~        & 0        & 0.967  & 0.167  &  0.992 &  0.971  & 0.573 & 0.814\\
POD     & ~~~0 $\rightarrow$ 1~        & 1        & 0.917  & 0.714  &  0.617 &  0.859  & 0.490 & 0.554\\
POFD    & ~~~0 $\rightarrow$ 1~        & 0        & \ldots & 0.021  &  0.305 &  0.119  & 0.002 & 0.055\\
TSS     & ~$-$1 $\rightarrow$ 1~       & 1        & \ldots & 0.693  &  0.312 &  0.740  & 0.488 & 0.500\\
HSS     & ~$-$1 $\rightarrow$ 1~       & 1        & 0.169  & 0.739  &  0.008 &  0.049  & 0.455 & 0.255\\
BIAS    & ~~~~0 $\rightarrow$ $\infty$ & 1        & \ldots & 0.857  & 80.156 & 29.413  & 1.147 & 2.973\\
\hline
\end{tabular}
\end{table*}

Compared to the previous metrics, POD and POFD more naturally illustrate the forecast success of MFWs. The POD tells us that MMCOs correctly forecast 55.4\% of days that contain flaring $\geqslant$\,X1 as MFWs, indicating moderate-to-good success in the observed $\geqslant$\,X1 days category. Meanwhile the POFD tells us that MMCOs incorrectly forecast just 5.5\% of days with flaring $<$\,X1 as MFWs, which is significantly low and indicates very good success in the observed $<$\,X1 days category. Together these two quantities result in MFWs achieving a reasonable TSS value of 0.500, exactly mid-way between a ``no skill'' forecast and a ``perfect'' forecast.

It is clear from their relative contributions to TSS that the category with most room to improve upon is that of POD ({\it i.e.}, better catching of $\geqslant$\,X1 days by MFWs) rather than that of POFD. This reaffirms the findings of the thought experiments in Section~\ref{ss:res_mfw_var} and the discussion of the source active regions of our missed flares in Section~\ref{ss:res_mis_fla}. It is worth recalling that, if a solar observing satellite had existed at an L5 vantage point, 11 of the 38 missed $\geqslant$\,X1 flare days might have been correctly forecast as MFWs from these regions located beyond or close to the Earth-viewed East limb. This hypothetical scenario would change TP and FN in Table~\ref{tbl:mfw_ct} to 52 and 22, respectively (because only the observed $\geqslant$\,X1 row would be affected), resulting in a very good POD of 0.703, the same POFD of 0.054 (since the observed $<$\,X1 row would be unaffected), and a very good TSS of 0.649.

\subsection{Comparison to Other Forecast Performance}\label{ss:oth_met_sss}

To place the performance of MFWs in proper context, it is necessary to compare the metrics discussed to those of other $\geqslant$\,X1 forecasting efforts. Table~\ref{tbl:mfw_sss} includes values, where available, from five published works (listed in chronological order):
\begin{itemize}
\item column 4 -- the neural network method of \citet{Colak:2009} applied to \citet{McIntosh:1990} sunspot classifications;
\item column 5 -- the ordinal logistic regression method of \citet{Song:2009} applied to total unsigned magnetic flux, length of strong-gradient neutral line, and total magnetic dissipation (specifically, model 4);
\item column 6 -- the superposed epoch analysis method of \citet{Mason:2010} applied to gradient-weighted inversion-line length;
\item column 7 -- the Poisson statistics method of \citet{Gallagher:2002} applied to McIntosh sunspot classifications, converted to categorical forecasts that optimize TSS in \citet{Bloomfield:2012};
\item column 8 -- the NOAA Space Weather Prediction Centre (SWPC) human-modified operational forecasts \citep{Crown:2012}.
\end{itemize}
As mentioned in Section~\ref{ss:ver_met_sss}, the preferred metric for intercomparison of forecasts from different time ranges is TSS due to its insensitivity to the underlying event climatology. It is unfortunate that TSS is unable to be calculated for the work of \citet{Colak:2009}, especially as this method achieves a very large POD of 0.917 compared to the 0.554 for MFWs. However, the very large FAR of 0.967 and low HSS of 0.169 (albeit climatology dependent) make it likely that their POFD would be quite large and result in, at best, a moderate TSS.

The \citet{Song:2009} TSS of 0.693 and FAR of 0.167 are among the best so far published for $\geqslant$\,X1 flares, but need to be considered susceptible to large error bars because that study included only seven flares and 55 forecasts. The study by \citet{Mason:2010} contains better number statistics (on the order of tens of thousands) and achieves a good POD of 0.617 that is better than for MFWs. However, this method produces a higher POFD of 0.35 that lead to a TSS of 0.312 (lower than for MFWs), which is likely in part due to their forecast interval being just 6\,h. In addition, having a BIAS of $\approx$\,80 and a FAR of 0.992 makes it extremely cost-inefficient to follow these forecasts in terms of the observing time that would have to be spent to catch even one $\geqslant$\,X1 day.

The largest TSS reported in Table~\ref{tbl:mfw_sss} is that of \citet{Bloomfield:2012}, which was found through varying a probability threshold and creating categorical forecasts to maximize TSS. It should be noted that this optimum TSS value was found for a threshold probability of 1\%, meaning that for this method X-class flares were \emph{always} forecast for any McIntosh classification that had historically produced any X-class activity ({\it i.e.}, in the previous two solar cycles from which the Poisson flare rates were determined) and \emph{never} forecast for any McIntosh classifications that historically produced no X-class activity. Although this perceived lack of skill does not preclude its usefulness for telescope/satellite observers, the BIAS of $\approx$\,30 and FAR of 0.971 would again lead to a substantial observing time inefficiency if these forecasts were followed.

Finally, the assessment of the NOAA/SWPC operational forecasts carried out by \citet{Crown:2012} shows several differences in performance from the MFWs. Similar to \citet{Bloomfield:2012}, \citet{Crown:2012} choose a probability threshold (25\%) for converting into categorical forecasts but do so by optimizing a different metric (the critical success index rather than TSS). By doing so they achieve the highest ACC (0.996), lowest FAR (0.573), near-best BIAS (0.147 away from unity; \citet{Song:2009} achieve 0.143 away from unity), and lowest POFD (0.002). However, their POD of 0.490 is slightly lower than for MFWs and the resulting TSS of 0.488 is almost identical to that found here for MFWs (0.500).

\section{Future Activities}\label{s:future}

After 31 May 2010, the MFW criteria were altered in anticipation of the expected weaker amplitude of solar cycle 24. At that time the Max Millennium program changed its definition of a major flare to $\geqslant$\,M5, affecting the magnitude of flares being targeted by subsequent MFWs and the first criterion in Section~\ref{ss:mfw_cri}. At the same time, sunspot area thresholds in the second to fourth criteria were relaxed from the previously stringent 500\,$\mu$H to subjective definitions of several-100\,$\mu$H. This was introduced to allow MMCOs more flexibility in deciding to call, or as importantly to not call, a MFW. However, this change to forecasting at a lower flaring level means that the forecast performance for MFWs after 31 May 2010 will not be readily comparable to the results presented here.

Further changes to the Max Millennium program involve two new Target of Opportunity (ToO) Observing Plans that have been put in place since 2014. The first addition is OP 018 \emph{Region Likely to Produce Great Flares} that aims to coordinate observations of: i) flares with nuclear-line or gamma-ray emission extending above 50\,MeV (typically high M-/X- class flares that are associated with fast and wide CMEs), or ii) flares $\geqslant$~X5, with or without such gamma-ray emission. This OP was designed (in part) to support the Fermi PI team, who incorporated the issuing of a Great Flare Watch into choosing whether or not to reorient the satellite for optimal solar observations by the Large Area Telescope \citep{Atwood:2009} at energies above 100\,MeV. The second addition triggers a ToO for the NSO Dunn Solar Telescope (DST) during their new form of Service Mode Operations -- designated as OP 020 \emph{DST Service Mode Support}. All DST observations over 1\,--\,31 October 2014 were performed in this service mode and OP 020 coordinated flare observations with RHESSI, IRIS, \emph{Hinode}, and specific instruments on SDO. The campaign was hugely successful, with all instruments catching flares from NOAA 12192 -- the largest sunspot group in 24\,years -- over a wide range of magnitudes ({\it i.e.}, B- to X- class). This MMCO activity reaffirmed the value that such campaigns bring to the solar flare community and endorses the use of the Max Millennium program to facilitate more coordinated efforts in the future. This will include further month-long service mode operations at the DST, as these are a test-bed for how observations at the Daniel K. Inoue Solar Telescope (DKIST) will be performed when it comes online in 2018.

The Max Millennium team occasionally considers adding quantitative active region properties to supplement the existing qualitative morphological properties in the MFW criteria of Section~\ref{ss:mfw_cri}. Several properties derived from line-of-sight (LOS) magnetograms show a correlation with flaring activity -- {\it e.g.}, $R$ \citep{Schrijver:2007}, $B_{\mathrm{eff}}$ \citep{Georgoulis:2007}, and $^{\mathrm{L}}$WL$_{\mathrm{SG}}$ \citep{Falconer:2008} -- and access to vector magnetograms from SDO has yielded promising results when coupled with machine-learning methods \citep{Bobra:2015}. However, more research is required before any such derived LOS or vector magnetogram quantities can be used as quantitive threshold criteria for MFWs.

\section{Conclusions}\label{s:conc}

In this paper we have reviewed the performance of the Max Millennium MFW alerts from 1 February 2001 to 31 May 2010, inclusive. The results presented here highlight a satisfying level of performance in terms of numbers and percentages of $\geqslant$\,X1 flares being caught by MFW target regions. In the time range considered, the 220 issued MFWs amount to just 6.5\% of the available days (3\,407). Although three uninterrupted MFW periods lasted for more than 15\,days, most of the 32 periods lasted 2\,--\,8 days. If observers reacted immediately to the first MFW in each uninterrupted period and followed these in full, they would have caught 56\% of the 87 $\geqslant$\,X1 flares that occurred in the whole time range. Interestingly, these occurred in just 19\% of the MFW days. Even with 24\,h delays in responding to the first MFW in a period (typical of delays due to daily satellite planning meetings and telecommanding schedules for scientific observations), the percentage of flares $\geqslant$\,X1 that would have been caught only decreases to 53\%.

In addition to these characteristics, the time scales on which MFW periods start and end were studied. There is strong evidence that MFW periods terminate at suitable times, through essentially no gain in major flares that could be caught in the 24\,h that follow MFWs ceasing and a loss of some X1\,--\,X3 flares if MFWs ended 24\,h earlier. In contrast, there is evidence that the start time of MFW periods could improve, through a gain in the $\geqslant$\,X1 flares caught if starting 24\,h earlier. This potential improvement highlights the ``holy grail'' of catching the first major flare from an active region. However, only 41\% of MFW periods ({\it i.e.}, 13 out of 32) were initiated by MMCOs in reaction to a major flare.

The real forecasting capabilities of MFWs were investigated using standard forecast verification metrics and skill scores. The MFWs achieve reasonable levels of performance with a TSS of 0.500, while a BIAS of $\approx$\,3 indicates only a minor level of overprediction compared to some of the published works. In particular, the MFW TSS is comparable to a categorical flare/no-flare interpretation of the NOAA/SWPC probabilistic forecasts (0.488). It is interesting to see that expert forecasting systems ({\it i.e.}, those involving human decisions) are still towards the top of the performance table when considering ACC, TSS, and BIAS.

In summary, the Max Millennium program has provided a successful avenue for coordinating observations of major solar flares through the MFW email alerts. The very small observing overhead of following MFW target regions represents an efficient use of ground- and space- based instruments attempting to study the physical processes triggering solar flares and governing their evolution.

%
\begin{acks}
The authors wish to thank the excellent efforts of both our newest MMCO (Ying Li) and past MMCOs who have contributed since 2001 (in alphabetical surname order: Paul A. Higgins, R. T. James McAteer, and Claire L. Raftery), Keiji Yoshimura for maintaining the Max Millennium website, and the referee for useful comments that helped improve the manuscript. The Max Millennium program has been supported by the enlightened RHESSI PI team led by Principle Investigator Robert P. Lin (later S\"{a}m Krucker) and Project Scientist Brian R. Dennis through Sub-agreement No. SA-1868 26308PG between University of California, Berkeley and Montana State University. DSB received funding from the European Space Agency PRODEX Programme and the European Union's Horizon 2020 Research and Innovation Programme under grant agreement No.~640216 (FLARECAST project). ROM received funding from NASA LWS/TR\&T grant NNX11AQ53G and NASA LWS/SDO Data Analysis grant NNX14AE07G.
\end{acks}

%
 \bibliographystyle{spr-mp-sola}
 \bibliography{ms.bib}

\end{article} 
\end{document}